**Omnidirectional photonic chiral flatband in nonlocal membrane metasurfaces**

Baohe Zhang,[1,†] Jumin Qiu,[2,3,†] Meng Qin,[1] Haiyan Jiang,[1] Shuyuan Xiao,[3,*] Zhiwei Zheng,[4,5] Leman Kuang,[4,5] Hui Jing,[4,5] Xinxing Zhou,[4,5,*]and Hongju Li[1,*]

[1]School of Physics, Hefei University of Technology, Hefei, Anhui 230009, China

[2]School of Physics and Materials Science, Nanchang University, Nanchang 330031, China

[3]School of Information Engineering, Nanchang University, Nanchang 330031, China

[4]Key Laboratory of Low-Dimensional Quantum Structures and Quantum Control of Ministry of Education, School of Physics and Electronics, Hunan Normal University, Changsha 410081, China

[5]Hunan Research Center of the Basic Discipline for Quantum Effects and Quantum Technologies, Hunan Normal University, Changsha 410081, China

[†]Baohe Zhang and Jumin Qiu contributed equally to this work.

[*]Correspondence: syxiao@ncu.edu.cn (S. X.); xinxingzhou@hunnu.edu.cn (X. Z.); hjli@hfut.edu.cn (H. L.)

## Abstract

Omnidirectional flat-band resonances, characterized by an enhanced photonic density of states and inherent angular robustness, are highly sought-after in integrated nanophotonic devices, particularly when integrated with chiral functionality. Here we realize such resonances in a nonlocal silicon membrane metasurface patterned with periodic square-lattice air-hole arrays. Increasing the lattice period not only compresses the Brillouin zone but, crucially, weakens the evanescent coupling between neighbouring Bloch modes associated with the same-order guided resonances. Driven by the tight-binding model in the limit of weak inter-unit-cell coupling, the pronounced band flattening of the degenerate guided resonance along both $k_x$ and $k_y$ yields, giving rise to an omnidirectional flat-band resonance. Remarkably, both numerical simulations and experiments reveal a universal route for endowing flat-band guided resonances with optical chirality through the deliberate breaking of the mirror symmetry of air holes. As a result, the omnidirectional chiral flat-band resonance emerges along both principal in-plane directions, with $Q$-factors exceeding $10^3$ and circular dichroism greater than 0.9 over a wide angular range of $\pm 5°$. Nonlinear measurements further show that the resulting resonance not only drives highly efficient third-harmonic generation but also imparts a pronounced spin-selective character to the nonlinear process. Simultaneously, the highly efficient nonlinear process also enables chirality-controlled frequency-upconversion imaging. Our results establish a general paradigm for engineering omnidirectional chiral flat-band resonances in planar silicon platforms, opening new opportunities for nonlinear nanophotonics and chiral imaging.

## Introduction

Flat bands, originally explored in condensed-matter systems in connection with correlated electronic phases,[1,2] have recently attracted growing attention in nanophotonics.[3] The optical flat band describe the dispersionless resonant eigenfrequency in the momentum space where the kinetic energy of photonics is quenched and the density of states is strongly enhanced.[4] Owing to the unconventional light localization and slow-light propagation, the light-matter interactions based on flat-band resonances have been substantially enhanced.[5-8] A flat band in momentum space further implies that, when excited from free space, the resonance wavelength and quality factor ($Q$-factor) becomes largely independent of the angle of incidence.[9] This intrinsic angular robustness is highly advantageous for device integration in nanophotonics.[10,11] Inspired by the realization of flat bands in electronic systems, the primary route to photonic flat bands likewise relies on precise lattice arrangement engineering.[12,13] Metasurfaces, comprised of artificial atom arrays of subwavelength nanostructures, therefore, emerge as an ideal platform for photonic flat bands owing to their exceptional ability to control both optical localization and radiation.[14-19] Canonical lattice geometries, including Lieb,[20-22] Kagome[23,24] and diamond lattices,[25] have been successfully implemented in metasurfaces. Through mechanisms such as destructive interference and the formation of compact localized states, spectrally resolved photonic flat bands have been experimentally observed. Beyond single-layer designs, twisted bilayer metasurfaces have enabled the emergence of moiré-induced flat bands at so-called magic angles,[26,27] while related displaced bilayer metagratings have also been shown to support flat-band formation.[28] However, achieving photonic flat bands in these systems typically requires highly precise and stringent design of the constituent meta-atoms, such that resonant modes are tightly confined at the single-unit-cell level and inter-lattice coupling remains weak. This stringent requirement on modal localization imposes considerable demands on nanofabrication accuracy, posing a significant challenge for practical implementation.

From the perspective of enhancing light-matter interactions, high $Q$-factor photonic flat bands offer a unique advantage.[29] They combine an exceptionally large photonic density of states in space with prolonged photon lifetimes in time. This dual enhancement enables the strengthening of light-matter coupling across both spatial and temporal dimensions. At the same time, achieving strong mode localization at the single-unit-cell level while maintaining robustness against fabrication imperfections is crucial for practical implementations. In this regard, planar metasurfaces and photonic crystal arrays provide an ideal platform for realizing high $Q$-factor photonic flat bands. Various localized states including anapoles,[4,17,18] bound states in the continuum,[30-34] and guided-mode resonances[35,36] have been proposed. Photonic flat bands based on these localized states can arise from two primary mechanisms: strong interband coupling induced by Brillouin zone folding,[37,38] and the tight-binding model with weak inter-lattice coupling, typically such as coupled-resonator optical waveguide (CROW) model.[39,40] These systems have exhibited remarkable performance across a range of applications, including electromagnetically induced transparency,[41] nonlinear optics,[29] thermal emission,[37] lasing[30] and integrated photodetection.[36] The integration of high $Q$-factor photonic flat bands with chirality—forming chiral photonic flat bands—offers a compelling route to simultaneously enhance light-matter interactions and expand the available degrees of freedom in photonic device control.[42,43] This emerging direction has attracted significant attention. However, achieving such states while maintaining high $Q$-factors, negligible spectral dispersion, and strong circular dichroism across a broad range of incident angles remains highly challenging. Recent studies have shown that anisotropic nonlocal metagratings and metasurfaces hold promise for realizing high $Q$-factor chiral photonic flat bands.[44-47] Yet, the demonstrated flat bands are typically limited to a specific wavevector direction in momentum space, rather than extending isotropically across the Brillouin zone. The realization of omnidirectional chiral flat bands with simultaneously high $Q$-factors remains an outstanding goal in the field.

In this work, we establish a general strategy for omnidirectional chiral flat-band resonances in nonlocal square-lattice membrane metasurfaces composed of periodically patterned air holes. Our analysis reveals that the evanescent coupling between neighbouring Bloch modes associated with same-order guided resonances progressively weakens as the lattice period increases. Governed by the tight-binding model in the limit of weak inter-mode coupling, the pronounced band flattening of the degenerate guided resonance along both $k_x$ and $k_y$ yields and thus an omnidirectional flat-band resonance emerges. By introducing in-plane symmetry breaking through tailored air-hole geometries, we further demonstrate a universal route to chiral flat-band guided resonances. Both numerical simulations and experimental measurements reveal a robust chiral flat-band guided resonance, exhibiting stable wavelength, $Q$-factors exceeding $10^3$, and circular dichroism above 0.9 within a wide-angle range of $\pm 5°$. Notably, such guided resonance is insensitive to the azimuthal angle, confirming its omnidirectional nature. Nonlinear optical measurements further suggest that this resonance efficiently drives third-harmonic generation (THG), achieving a conversion efficiency of $1.09\times10^{-6}$ at a pump intensity of 6.4 GW cm$^{-2}$. Importantly, the generated THG signal inherits the chiral characteristic of the underlying resonance, exhibiting a polarization contrast of $4.04\times10^3$ between opposite circular polarizations. This enhanced THG further enables chirality-resolved nonlinear imaging with a spatial resolution of ~3.2 μm. Together, our results establish a general paradigm for realizing omnidirectional chiral flatbands in nonlocal membrane metasurfaces. The observed omnidirectional chiral flat-band guided resonance opens new opportunities for nonlinear nanophotonics and chiral imaging.

## Results

### Working principle and numerical simulations

It is well known that an infinite isotropic silicon (Si) slab supports a hierarchy of transverse-electric (TE) and transverse-magnetic (TM) guided modes, whose electric fields are predominantly polarized in plane and out of plane, respectively. Introducing a square-lattice periodic modulation transforms these slab-guided modes into Bloch guided resonances. The resulting periodic potential folds the continuous slab dispersion into the first Brillouin zone, where Bragg scattering lifts the degeneracies of the guided continuum and reconstructs it into discrete photonic bands with distinct symmetries and radiative characteristics. We investigate the intrinsic guided resonances arising in a Si film of thickness $H$ = 220 nm supported on a silica ($SiO_2$) substrate, where an in-plane periodic modulation transforms the slab-guided modes into Bloch resonances with engineered dispersive properties. The band structure of the resulting guided resonances is calculated using the eigenmode solver in COMSOL Multiphysics. Floquet periodic boundary conditions are applied along the $x$ and $y$ directions to define the in-plane wavevectors $k_x$ and $k_y$, while perfectly matched layers are implemented along the $z$ direction to emulate an open, reflection-free system. In the near-infrared regime, both Si and $SiO_2$ are treated as lossless dielectrics, with refractive indices of 3.48 and 1.45, respectively. Here we focus on a TM-derived guided resonance supported by a square-lattice photonic crystal with lattice period $P$ = 900 nm, whose field profile forms a second-order standing-wave resonance along both the $x$ and $y$ directions. As shown in the bottom panel of Fig. 1(b), the out-of-plane electric field $E_z$ exhibits identical spatial distributions along the two in-plane directions, implying degenerate dispersions along $k_x$ and $k_y$. For conceptual clarity and to distinguish different orders of standing-wave resonances within a single unit cell, we adopt a mode-labelling scheme in which the first subscript denotes the number of nodes of the out-of-plane electric field $E_z$ along the $x$ direction, while the second corresponds to that along $y$. Accordingly, the guided resonance investigated here can be identified as a $TM_{22}$ guided resonance. Physically, the mode may be approximately interpreted as a coherent superposition of two orthogonal standing-wave resonances, $TM_{20}$ and $TM_{02}$, as illustrated in Fig. 1(b). As the lattice period $P$ increases, two effects emerge simultaneously. First, the first Brillouin zone contracts, causing the slab dispersion to fold into a progressively smaller momentum space. More importantly, the modal field becomes increasingly delocalized within each unit cell, thereby reducing the evanescent coupling between neighbouring Bloch modes associated with the guided resonances. In the weak-coupling regime, this behaviour is naturally captured by a tight-binding picture, in which the dispersion of the $TM_{22}$ guided resonance becomes progressively flatter (see details in Section 1 of Supporting Information).[48,49] For lattice periods exceeding $P$ = 900 nm, a pronounced flat-band guided resonance emerges, as shown in upper panel of Fig. 1(c). Owing to the isotropic symmetry of the $TM_{22}$ mode, the resulting flat band is intrinsically omnidirectional in momentum space, particularly along the $k_x$ and $k_y$ directions.

Meanwhile, the guided mode of the Si membrane is intrinsically decoupled from free-space radiation, resulting in vanishing radiative losses and an ultrahigh $Q$-factor exceeding $10^9$, as shown in bottom panel of Fig. 1(c). These results establish the square-lattice-modulated Si slab as a powerful platform for realizing ultrahigh-$Q$-factor omnidirectional flat-band resonances.

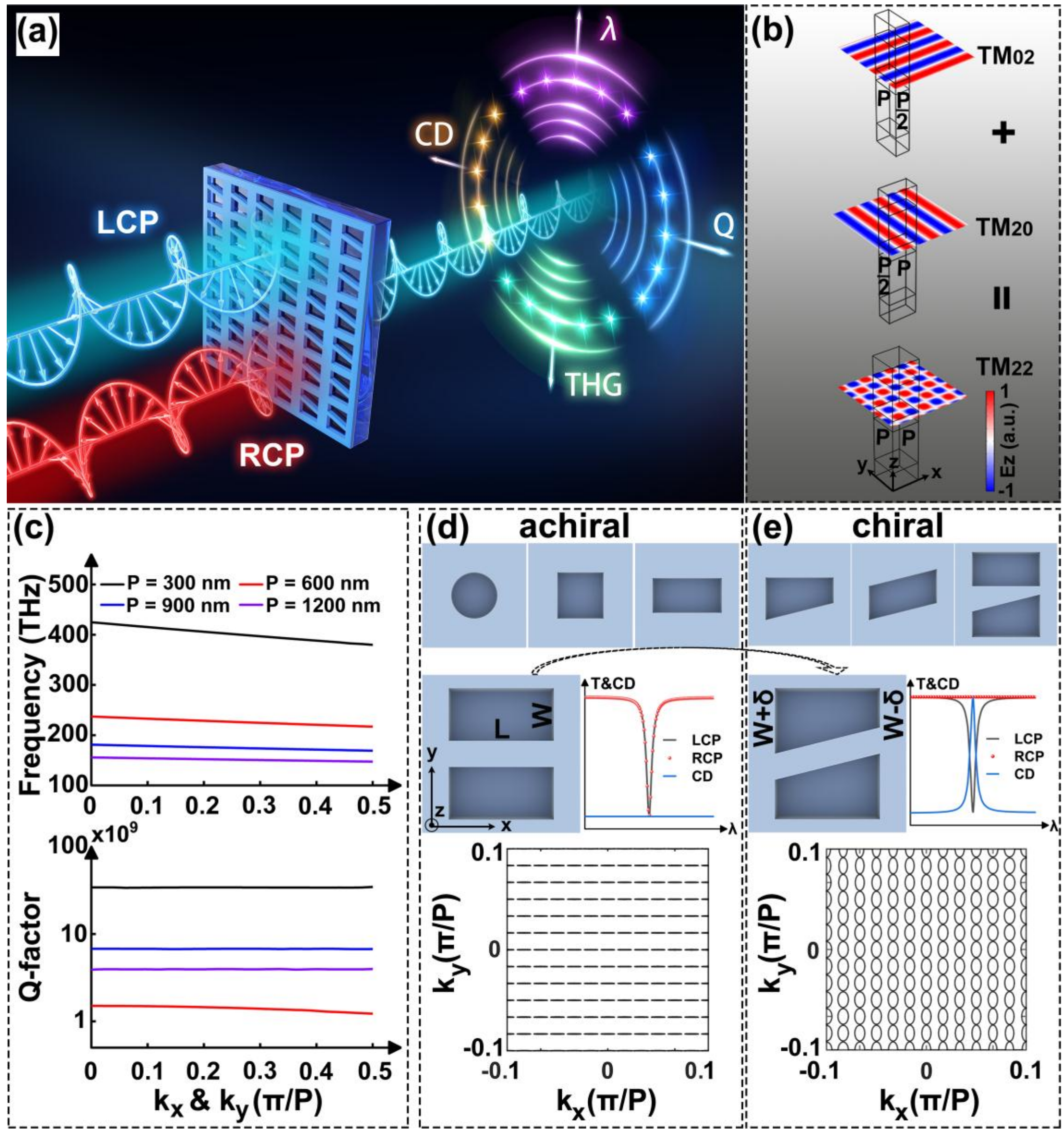

**Fig. 1. Concept of omnidirectional chiral flatband in nonlocal membrane metasurfaces.** (a) Conceptual illustration showing that the omnidirectional chiral flat-band guided resonance supported by the nonlocal membrane metasurface with the unit cell consisting of double-trapezoidal holes, simultaneously exhibiting angle-independent resonance wavelength $\lambda$, $Q$-factor ($Q$), circular dichroism (CD) and third-harmonic generation (THG) response. (b) Schematic of guided resonances in a silicon slab with in-plane periodic modulation. The $TM_{22}$ guided resonance can be approximately understood as a coherent superposition of two orthogonal second-order standing-wave resonances, $TM_{20}$ and $TM_{02}$, within a unit cell. Rectangular lattices with $P_x = P/2$, $P_y = P$ and $P_x = P$, $P_y = P/2$ selectively support the $TM_{02}$ and $TM_{20}$ resonances, respectively, whereas a square lattice with $P_x = P_y = P$ supports the isotropic $TM_{22}$ guided resonance. (c) Band dispersion and corresponding $Q$-factor distribution of the $TM_{22}$ guided resonance along the $k_x$ and $k_y$ directions, as a function of the square lattice period $P$. (d) The top panel schematically show that introducing air holes with mirror-symmetric geometry transforms the slab-guided mode into an achiral guided resonance. The bottom panel shows the far-field polarization distribution of the $TM_{22}$ guided resonance in a nonlocal membrane metasurface with double-rectangular holes in each unit cell. (e) The top panel schematically shows that breaking the in-plane mirror symmetry of the hole geometry converts the slab-guided mode into a chiral guided resonance. The bottom panel shows the far-field polarization distribution of the $TM_{22}$ guided resonance in a nonlocal membrane metasurface with the unit cell consisting of double-trapezoidal holes.

To render slab-guided modes spectrally accessible as guided resonances, a natural strategy is to introduce a periodic array of air holes, thereby breaking the continuous translational symmetry of the Si slab and simultaneously transforming the Si film into a nonlocal membrane metasurface. Intriguingly, for the $TM_{22}$ guided resonance, the equivalent field distributions along the $x$ and $y$ directions endow the mode with both $C_2$ rotational symmetry and inversion symmetry, as shown in bottom panel of Fig. 1(b). As a consequence, the geometric symmetry of the air holes becomes the determining factor governing the emergence of chirality in the guided resonance. When the introduced holes preserve in-plane mirror symmetry—such as circular, square or rectangular shapes—the resulting guided resonance remains achiral, as shown in Fig. 1(d) (see details in Section 2 of Supporting Information). In contrast, breaking the in-plane mirror symmetry of the are holes—such as parallelogram or right-angled trapezoidal shapes—leads to the emergence of chiral guided resonances, as illustrated in Fig. 1(e). To elucidate this effect, we calculate the momentum-space far-field polarization states of the $TM_{22}$ guided resonance in the membrane metasurface with the unit cell consisting of double-rectangular holes and their symmetry-broken counterparts (double right-angled trapezoidal holes), respectively. For the mirror-symmetric double-rectangular geometry, the far-field polarization remains linear throughout the first

Brillouin zone, indicating an achiral response, as shown in the bottom panel of Fig. 1(d). By contrast, the symmetry-broken double-trapezoidal geometry produces circularly polarized emission across the entire first Brillouin zone, confirming the emergence of chirality, as demonstrated in the bottom panel of Fig. 1(e). Importantly, as long as the lattice period remains unchanged, the excited resonance retains the omnidirectional flat-band characteristic of the $TM_{22}$ guided resonance, irrespective of the specific air-hole geometry. This establishes a general route to omnidirectional chiral flat-band resonances in nonlocal Si membrane metasurfaces through the introduction of in-plane mirror-symmetry breaking in the air-hole array. By harnessing these resonances to drive third-harmonic generation (THG), we will realize a remarkably robust nonlinear response in which key observables—including the resonance wavelength ($\lambda$), $Q$-factor ($Q$), circular dichroism (CD), and THG efficiency—remain essentially insensitive to the angle of incidence, as schematically illustrated in Fig. 1(a).

To explicitly elucidate the spectral evolution underlying the emergence of chirality in the $TM_{22}$ guided resonance, we consider the nonlocal membrane metasurface with unit cell composed of double right-angled trapezoidal holes. As illustrated in Figs. 1(d) and 1(e), this geometry is derived from a pair of identical rectangular holes. The initial rectangular dimer, with length $L$ and width $W$, is symmetrically arranged within the unit cell, preserving both in-plane $C_2$ rotational symmetry and mirror symmetry. By increasing the width of one side of each rectangle by $\delta$ while decreasing the opposite side by the same amount, the rectangles are transformed into right-angled trapezoids. This deformation preserves the $C_2$ symmetry but breaks the in-plane mirror symmetry, thereby enabling the possibility of transition from an achiral $TM_{22}$ guided resonance to a chiral guided resonance. In our analysis, we set $L$ = 600 nm and $W$ = 300 nm, and define an asymmetry parameter $\alpha = \delta/W$. The dependence of the transmission spectra on $\alpha$ under right- (RCP) and left-circularly polarized (LCP) excitation is shown in Figs. 2(a) and 2(b), respectively. A direct comparison between Figs. 2(a) and 2(b) reveals that, as the asymmetry parameter $\alpha$ increases, the $TM_{22}$ guided resonance evolves into a spectrally accessible transmission dip with an increased linewidth and a pronounced helicity-dependent response to circularly polarized light. This behaviour confirms that breaking in-plane mirror symmetry effectively converts the $TM_{22}$ guided resonance into a chiral resonance. To quantitatively characterize the chiral response, we define the CD based on Jones matrix of transmission coefficient under circular polarization basis. The transmission coefficients in the circular polarization basis are expressed as[50-52]

$$J_{circ} = \begin{pmatrix} t_{rr} & t_{rl} \\ t_{lr} & t_{ll} \end{pmatrix}. \tag{1}$$

Here, the subscripts $r$ and $l$ represent the RCP and LCP, respectively. The $t_{ij}$ ($i = r$, $l$ and $j = r$, $l$) stands for the transmission coefficient of output polarization $i$ from input polarization $j$. The transmission of the incident RCP and LCP lights can be respectively expressed as

$$\begin{aligned} T_{RCP} &= |t_{rr}|^2 + |t_{lr}|^2 = T_{rr} + T_{lr}, \\ T_{LCP} &= |t_{ll}|^2 + |t_{rl}|^2 = T_{ll} + T_{rl}. \end{aligned} \tag{2}$$

The $T_{ij} = |t_{ij}|^2$ is the transmission of the membrane metasurfaces. The CD, defined as the normalized transmission difference, thus is described as[53-56]

$$\mathrm{CD} = \frac{(T_{rr} + T_{lr}) - (T_{ll} + T_{rl})}{(T_{rr} + T_{lr}) + (T_{ll} + T_{rl})}. \tag{3}$$

**Fig. 2. Numerical analysis of omnidirectional chiral flatband.** (a) and (b) correspond to transmission spectra as a function of the asymmetry parameter $\alpha$ under the RCP and LCP excitation, respectively. (c) The evolution of CD spectra on the asymmetry parameter $\alpha$. (d) $Q$-factor of the $TM_{22}$ guided resonance versus the asymmetry parameter $\alpha$. (e) Transmission spectra under circularly polarized excitation together with the corresponding CD spectrum for the membrane metasurface with $\alpha = 0.3$. (f) Transmission spectra of the four Jones matrix elements in the circular polarization basis at $\alpha = 0.3$. (g)-(i) Angle-resolved optical response of the $TM_{22}$ guided resonance under the oblique incidence along $x$ axis. Evolution of the resonant wavelength (g), $Q$-factor (h), and CD (i) as a function of the incidence angle ($\theta$) for the metasurfaces with different asymmetry parameters. (j)-(l) Azimuthal dependence of the chiral guided resonance at oblique incidence. Evolution of the resonant wavelength (j), $Q$-factor (k), and CD (l) as a function of the azimuthal angle ($\varphi$), for the metasurface of $\alpha = 0.3$ under an incidence angle of 5°.

The calculated CD spectra as a function of $\alpha$ are shown in Fig. 2(c). As anticipated, once the asymmetry parameter $\alpha$ deviates from zero—thereby breaking the in-plane mirror symmetry—a pronounced chiral resonance with CD exceeding 0.9 emerges. As $\alpha$ increases, the degree of symmetry breaking is enhanced, leading to stronger radiative leakage of the guided resonance and, consequently, a broader spectral linewidth. The extracted dependence of the $Q$-factor on the $\alpha$ is presented in Fig. 2(d), revealing a clear inverse quadratic relationship $Q \propto \alpha^{-2}$, a hallmark signature of the evolution of a leaky guided resonance. Furthermore, the sign of $\alpha$ directly determines the handedness of the CD. To elucidate the physical origin of the chirality, we take $\alpha = 0.3$ as a representative case. Figures 2(e) and 2(f) display the transmission spectra under LCP and RCP illumination, as well as the four transmission coefficients of the Jones matrix in the circular polarization basis, respectively. At $\alpha = 0.3$, a chiral guided resonance, with a $Q$-factor exceeding $10^3$ and a CD of 0.91, is observed at the wavelength of $\lambda = 1396$ nm. As shown in Fig. 2(f), this resonance is dominated by the imbalance between the co-polarized transmission channels $T_{rr}$ and $T_{ll}$, evidencing an intrinsically "true" chirality rather than one arising from polarization conversion. This chiral response originates from interference between co- and cross-polarized components in the linear polarization basis: constructive interference enhances the $T_{ll}$ channel, whereas destructive interference suppresses $T_{rr}$, ultimately giving rise to the observed strong chiral response (see details in Section 3 of Supporting Information). We further examine the angle-resolved response of the guided resonance. Figures 2(g)-2(i) present the dependence of the resonance wavelength, $Q$-factor, and CD on the angle of incidence for different $\alpha$, respectively. Strikingly, for a fixed $\alpha$, all three quantities remain nearly invariant with respect to the incident angle, revealing the emergence of a chiral flat-band guided resonance, in excellent agreement with the band structure shown in Fig. 1(c). Crucially, owing to the symmetry-equivalent field distribution and degenerate dispersion of the $TM_{22}$ guided resonance along the $x$ and $y$ directions, oblique incidence along either axis yields an identical chiral flat-band response. This establishes the realization of an omnidirectional chiral flat-band resonance. To further elucidate the omnidirectional nature of this chiral flat-band resonance, we investigate the dependence of the chiral response on the azimuthal angle ($\varphi$) of oblique incidence for the metasurface with $\alpha = 0.3$ under a fixed incident angle of 5 °. The $\varphi = 0$ °and 90 ° correspond to oblique incidence along the $x$- and $y$-directions, respectively. Figures 2(j)-2(l) present the resonance wavelength, $Q$-factor, and CD of the chiral $TM_{22}$ guided resonance as functions of the azimuthal angle $\varphi$, respectively. Remarkably, all three quantities remain nearly invariant with respect to the azimuthal angle, providing compelling evidence for the realization of an omnidirectional chiral flat-band resonance in the nonlocal membrane metasurface.

**Experimental demonstration**

To experimentally validate the emergence of the chiral flat-band guided resonance, we fabricate a series of nonlocal membrane metasurfaces with the asymmetry parameter $\alpha$ varied from 0 to 0.5 in increments of 0.1 on a Si-on-insulator platform, comprising a 220-nm-thick top Si layer and a 2-μm buried oxide layer. The metasurface patterns are defined by electron-beam lithography, followed by metal deposition, lift-off, and dry etching. Full fabrication details are provided in Section 4 of Supporting Information. The representative scanning electron microscopy (SEM) image of the fabricated membrane metasurfaces with $\alpha = 0.3$ is presented in Fig. 3(b). The linear optical response of the fabricated metasurfaces is characterized using a custom-built microspectroscopy platform. Illumination is provided by a broadband picosecond supercontinuum source (SC-PRO, YSL Photonics). The incident beam is prepared by sequential passage through a linear polarizer, a quarter-wave plate, and a focusing objective before interacting with the metasurface. The transmitted light is then collected and analysed using an optical spectrum analyser (AQ6370D, Yokogawa). A schematic of the linear optical measurement setup is shown in Fig. 3(a), with further experimental details provided in the Section 5 of Supporting Information.

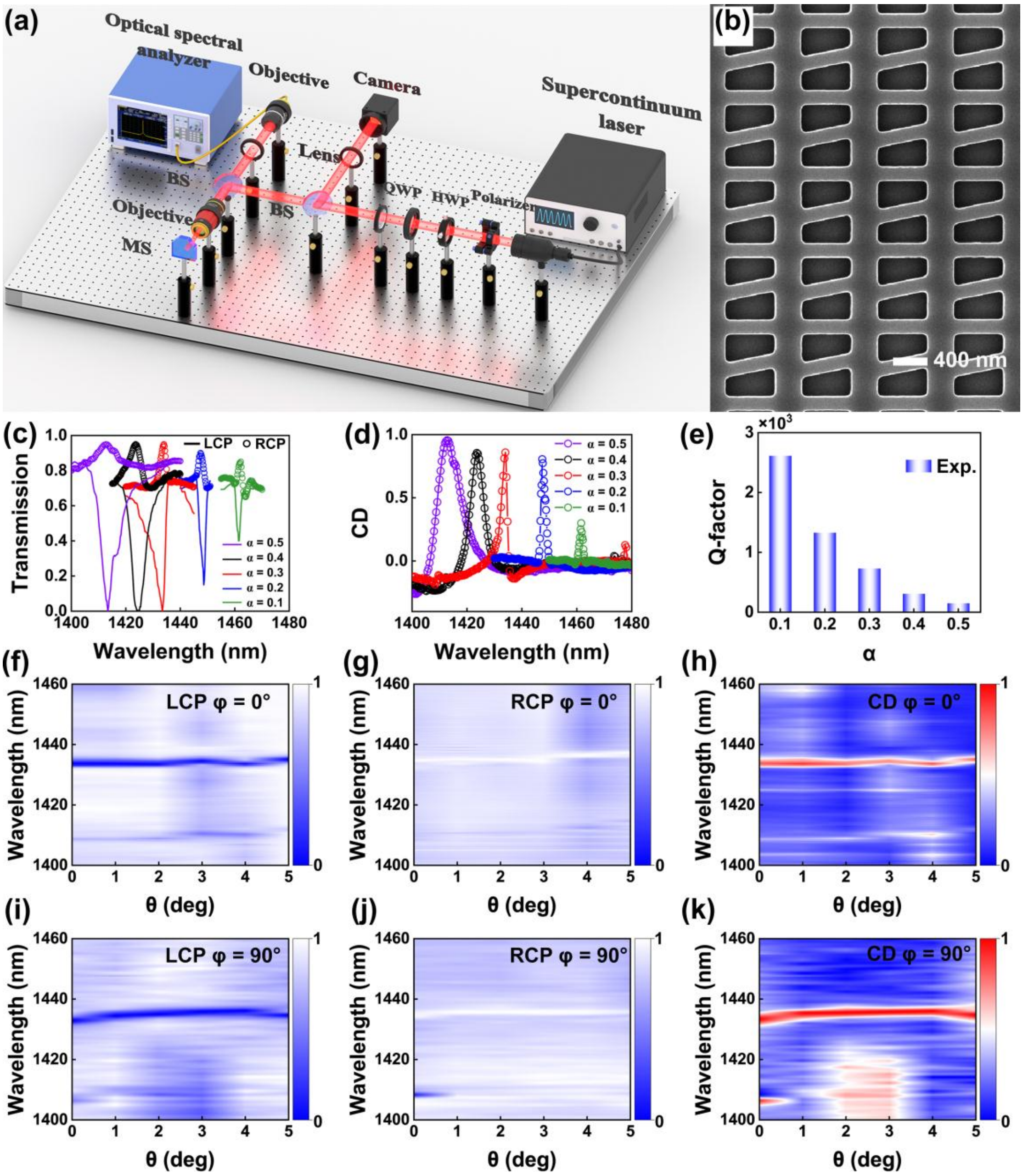

**Fig. 3. Experimental demonstration of omnidirectional chiral flatband.** (a) Schematic of the custom-built optical setup for linear transmission measurements. (b) SEM image of the nonlocal membrane metasurfaces with asymmetry parameter $\alpha = 0.3$. (c) Experimentally measured transmission spectra for the metasurfaces with different $\alpha$ under RCP and LCP illumination. (d) The experimentally measured CD spectra for the metasurfaces with different $\alpha$. (e) Experimentally extracted $Q$-factor of the $TM_{22}$ guided resonance as a function of the asymmetry parameter $\alpha$. (f)-(h) respectively correspond

to the experimentally measured angle-resolved LCP transmission spectra, RCP transmission spectra, and CD spectra of the metasurface with $\alpha$ = 0.3 under oblique incidence with an azimuthal angle of $\varphi$ = 0°. (i)-(k) respectively correspond to the experimentally measured angle-resolved LCP transmission spectra, RCP transmission spectra, and CD spectra of the metasurface with $\alpha$ = 0.3 under oblique incidence with an azimuthal angle of $\varphi$ = 90°. The $\varphi$ = 0° and 90° correspond to oblique incidence along the $x$- and $y$-directions, respectively.

The experimentally measured transmission spectra of the fabricated metasurfaces under normal incidence of LCP and RCP light are shown in Fig. 3(c), with the corresponding CD spectra presented in Fig. 3(d). In close agreement with numerical simulations, once the asymmetry parameter $\alpha$ deviates from zero, the spectral responses under LCP and RCP excitation become distinct, signifying the excitation of a chiral guided resonance. As $\alpha$ increases, symmetry breaking enhances the radiative losses of the guided resonance, resulting in spectral broadening and a corresponding reduction in the $Q$-factor. The experimentally extracted $Q$-factors faithfully follow the inverse-square scaling, $Q \propto a^{-2}$, as shown in Fig. 3(e), consistent with the characteristic evolution from a leaky guided resonance. Notably, for asymmetry parameters exceeding $\alpha$ = 0.2, we experimentally realize chiral guided resonances that simultaneously exhibit $Q$-factors above $10^3$ and CD greater than 0.9. To further experimentally verify that the chiral guided resonance inherits the flat-band character, we perform angle-resolved transmission measurements on the metasurface with $\alpha$ = 0.3. The measured transmission spectra under oblique RCP and LCP illumination along the $x$ direction, together with the corresponding CD spectra, are shown in Figs. 3(f)-3(h). Taken together, these measurements reveal an exceptionally robust resonance over an angular range of ±5°. Across this range, the resonance wavelength remains pinned at 1434 nm, while the $Q$-factor and CD are maintained at 750 and 0.91, respectively. This remarkable angular invariance provides compelling experimental evidence for the realization of a chiral flat-band guided resonance. Meanwhile, the Figs. 3(i)-3(k) respectively correspond to the experimentally measured angle-resolved LCP transmission spectra, RCP transmission spectra, and CD spectra of the metasurface under oblique incidence along the $y$-direction. Remarkably, a robust chiral flat-band resonance is likewise observed near $\lambda$ = 1434 nm. Therefore, the combined results presented in Figs. 3(f)-3(k) reveal highly robust chiral flat-band resonances simultaneously sustained along the $x$ and $y$ directions in the proposed nonlocal membrane metasurface, thereby providing experimental evidence for an omnidirectional chiral flat-band resonance.

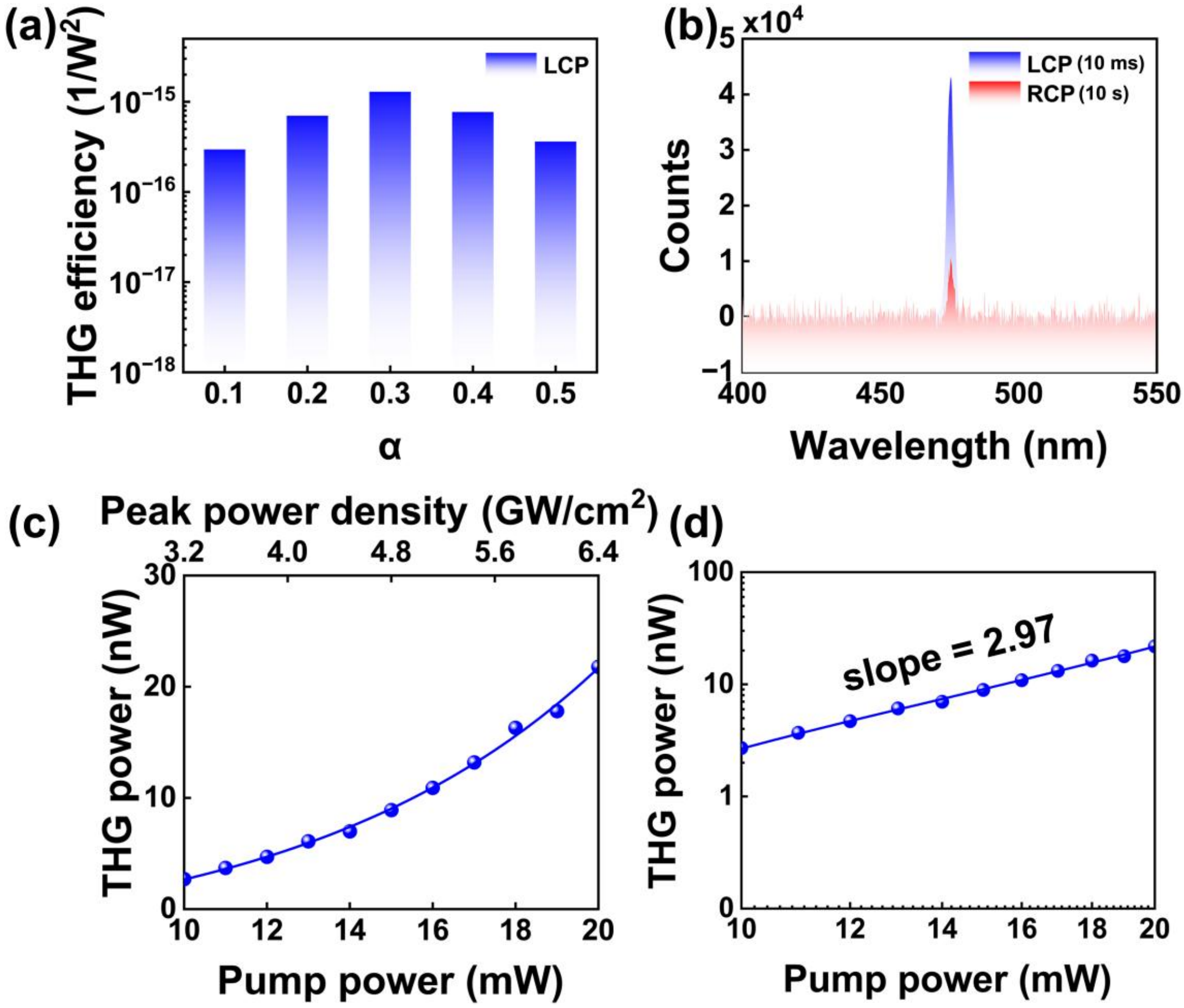

**Fig. 4. Experimental demonstration of chirality-controlled flat-band THG process.** (a) Normalized THG conversion efficiency as a function of the asymmetry parameter $\alpha$ under the LCP excitation at normal incidence. (b) Experimentally measured THG spectra from the nonlocal membrane metasurface with $\alpha$ = 0.3 under the LCP and RCP excitation, respectively. (c) Measured THG power as a function of the pump power for the membrane metasurface with $\alpha$ = 0.3 under LCP excitation. (d) The same data in (c) is plotted on logarithmic scales.

We next turn to the third-order nonlinear optical processes driven by the chiral flat-band guided resonance in the nonlocal Si membrane metasurface. The THG spectra and output power are measured using a custom-built nonlinear optical microscopy platform, schematically illustrated in the Section 5 of Supporting Information. Excitation is provided by a femtosecond fibre laser coupled to an optical parametric amplifier, delivering ultrashort, high-intensity pulses ideally suited for nonlinear spectroscopy. The pump wavelength is tuned to match the chiral guided resonance of each sample over the spectral range from 1410 to 1470 nm, with a repetition rate of 200 kHz and a pulse duration of approximately 200 fs. To quantify the intrinsic nonlinear enhancement afforded by the guided resonance, independent of excitation conditions, we define a normalized THG conversion efficiency as $\eta_{norm} = P_{THG}^{peak} / (P_{Pump}^{peak})^3$, where $P_{THG}^{peak}$ and $P_{Punp}^{peak}$ denote the peak powers of the emitted third-harmonic signal and the incident fundamental pump, respectively.[57-62] Because the chiral guided resonance is selectively excited only by LCP light, we first investigate the dependence of the normalized THG efficiency on the asymmetry parameter *α* under LCP excitation. The results are presented in Fig. 4(a). For any non-zero *α*, the chiral guided resonance is efficiently excited in the metasurface by the LCP pumping. Simultaneously, the field distribution is only weakly affected by variations in *α*, and thus the resonance-enhanced THG efficiency remains nearly unchanged on the order of $10^{-16}$ $W^{-2}$ for different *α*. By contrast, under RCP pumping, the guided resonance is largely inaccessible, and the THG efficiency is consequently strongly suppressed. As shown in Fig. 4(b), for the metasurface with *α* = 0.3, the THG intensity at *λ* = 478 nm is enhanced by a factor of $4.04 \times 10^3$ under LCP pumping relative to RCP pumping. This giant nonlinear contrast provides compelling evidence of chirality-controlled THG mediated by the chiral guided resonance. Furthermore, we experimentally measure the THG output power as a function of the pump power. As shown in Fig. 4(c), the THG power increases steeply with the pump power, as expected for a third-order nonlinear process. The power-law fitting of $P_{THG}^{average} = a(P_{Pump}^{average})^b$ yields a scaling exponent of $b = 2.97$, in excellent agreement with the cubic scaling of THG. The log-log plot shown in Fig. 4(d) further confirms this behaviour. As the average pump power is increased from 10 mW to 20 mW, the peak pump intensities of 3.2 to 6.4 GW $cm^{-2}$ are estimated from the pulse duration, repetition rate, and beam spot size of the pump light. The maximum average THG power reaches 21.8 nW at an average pump power of 20 mW. The absolute THG conversion efficiency, defined as $\eta = P_{THG}^{average} / P_{Punp}^{average}$, reaches $1.09 \times 10^{-6}$ at a peak pump intensity of 6.4 GW $cm^{-2}$, which is on the same order of magnitude as most previously reported Si-based metasurface systems.[63-66] It is worth noting that no saturation of the THG signal is observed within the accessible pump range, indicating that nonlinear loss channels, such as two-photon absorption and free-carrier absorption, remain negligible under our present operating conditions.

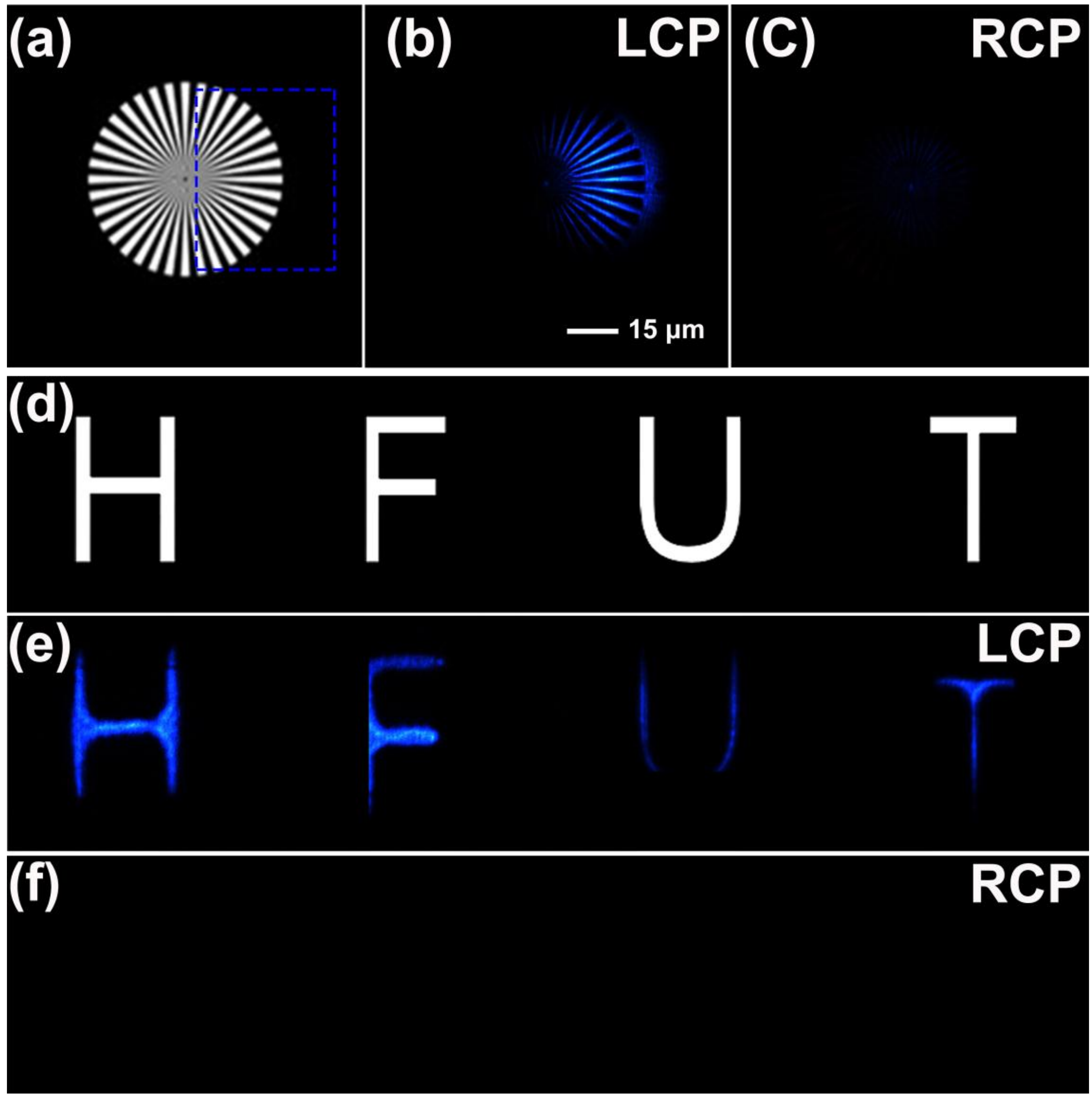

**Fig. 5. Experimental demonstration of chirality-controlled nonlinear frequency upconversion image.** (a) The Siemens star resolution target. (b) and (c) correspond to the upconverted visible images based on the Siemens star resolution target under the LCP and RCP pumping, respectively. (d) The letters of "HFUT" used for testing the upconverted visible image. (e) and (f) correspond to the upconverted visible images based on the target of "HFUT" under the LCP and RCP pumping, respectively.

Finally, we experimentally demonstrate the utility of the fabricated membrane metasurface for chirality-controlled nonlinear frequency upconversion imaging,[67-75] using the sample with $\alpha$ = 0.3 as a representative example. To maximize the THG efficiency, the pump wavelength is tuned to coincide with the resonant wavelength of $\lambda$ =1434 nm of the supported chiral guided resonance. A lens assembly expands the incident beam, enabling uniform illumination over a large area of the metasurface. In this configuration, an infrared test target projected onto the metasurface is upconverted to the visible range via THG process and subsequently recorded by an sCMOS camera. Further details of the imaging setup are provided in in the Section 5 of Supporting Information. We first employed a Siemens star as a standard resolution target, as shown in Fig. 5(a). Owing to its radially converging features, this target provides a stringent test of spatial resolution. Under LCP excitation, the chiral guided resonance is efficiently excited, leading to strong resonantly enhanced THG at the wavelength of $\lambda$ = 478 nm. As shown in Fig. 5(b), the overlapping region is clearly rendered in the visible, with the blue emission corresponding to the THG signal. More than a dozen concentric line pairs are distinctly resolved, corresponding to a spatial resolution of approximately 3.2 μm. By contrast, under RCP excitation, coupling to the chiral guided resonance is strongly suppressed. The resulting THG signal is too weak to support visible upconversion imaging, and no discernible image is observed, as shown in Fig. 5(c). To further demonstrate the generality of this approach, we image an arbitrary object, for example the letters "HFUT" (abbreviation for Hefei University of Technology). As shown in Figs. 5(d)-(f), the letters are clearly reconstructed under LCP illumination, enabled by the efficient THG process. In stark contrast, under RCP excitation, the resonance is effectively inaccessible, and the upconverted image vanishes. These nonlinear measured results establish that chiral flat-band guided resonances can serve not only as a platform for highly efficient, wide-angle, chirality-selective THG, but also as a powerful route to chirality-controlled nonlinear frequency-upconversion imaging.

## Discussion

In summary, we have demonstrated, both numerically and experimentally, a general strategy for omnidirectional chiral flat-band resonances in nonlocal Si membrane metasurfaces formed by a square lattice of air holes with broken in-plane mirror symmetry. Our analysis has revealed that introducing a square-lattice modulation transforms the slab-guided modes into guided resonances with degenerate dispersions along the $k_x$ and $k_y$ directions. Based on the tight-binding model, increasing the lattice period progressively suppresses inter-mode coupling and drives pronounced band flattening of the degenerate guided resonances, ultimately giving rise to an omnidirectional flat-band guided resonance. By further breaking the in-plane mirror symmetry of the air-hole geometry, the omnidirectional flat-band resonance acquires strong optical chirality. Numerical simulations and experimental measurements are in excellent agreement. Under oblique incidence along either the *x* or *y* direction, the wavelength of the resulting chiral flat-band guided resonance remains nearly invariant within $\pm 5°$, while maintaining *Q*-factors exceeding $10^3$ and CD above 0.9. Nonlinear measurements further show that the chiral flat-band guided resonance not only enables highly efficient THG through enhanced light confinement, but also endows the nonlinear process with pronounced spin selectivity. In particular, THG reaches a conversion efficiency of $1.09\times10^{-6}$ at a pump intensity of 6.4 GW cm$^{-2}$, and simultaneously exhibits an extraordinary polarization contrast up to $4.04\times10^3$ between opposite circular polarizations. Moreover, this highly efficient nonlinear interaction enables chirality-controlled frequency-upconversion imaging with a spatial resolution of approximately 3.2 μm. Our work establishes a versatile paradigm for engineering omnidirectional chiral flat-band resonances in planar dielectric metasurfaces, opening new opportunities for nonlinear nanophotonics, chiral light-matter interactions, and polarization-selective imaging.

## Methods

### Numerical simulations

All numerical simulations presented in this work are performed using the finite-element-method (FEM) solver implemented in COMSOL Multiphysics. The photonic band structures are calculated using the eigenfrequency solver. To determine the dispersion relation as a function of the in-plane wavevector, Floquet periodic boundary conditions are imposed along the *x* and *y* directions, while perfectly matched layers (PMLs) are employed along the out-of-plane *z* direction to absorb outgoing radiation and emulate an open photonic environment. The refractive indices of Si and $SiO_2$ are fixed at 3.48 and 1.45, respectively, throughout all simulations. The transmission spectra are obtained using the frequency-domain port model. The incident right- and left-handed circularly polarized (RCP and LCP) waves are defined by the Jones vectors $1/\sqrt{2}(1,i,0)$ and $1/\sqrt{2}(1,-i,0)$, respectively.

### Sample fabrication

The nonlocal Si membrane metasurfaces are fabricated on a Si-on-insulator (SOI) platform consisting of a 220-nm-thick silicon device layer and a 2-μm-thick buried oxide layer. Initially, the SOI wafer is sequentially cleaned in an ultrasonic bath using acetone, isopropyl alcohol, and deionized water, each for 10 minutes. After standard solvent cleaning, a 110-nm-thick layer of positive electron beam resist (ZEP-520A, Zeon Chemicals) is then spin-coated onto the cleaned wafer and prebaked on a hotplate at 180 °C for 3 minutes. The metasurface patterns, covering an area of $400 \times 400$ μm², are defined by high-resolution electron-beam lithography (100 keV) and subsequently transferred into the silicon layer via inductively coupled plasma etching. The residual resist is finally removed using N-methyl-2-pyrrolidone. Following resist removal, large-area nonlocal membrane metasurfaces with precisely controlled structural asymmetry are obtained. A schematic illustration of the fabrication workflow is provided in Fig. S4. Representative scanning electron microscopy image of the fabricated metasurfaces with the asymmetry parameter of $\alpha = 0.3$ is shown in Fig. 3(b).

**Linear optical characterization.**

The spectrally observed guided resonances are characterized using a home-built microscopic spectroscopy system based on cross-polarization measurements. The schematic of the homemade setup for linear spectral measurement is shown in Fig. 3(a). The picosecond pulsed laser (SC-PRO, YSL Photonics) is used as a light source to provide a broad spectrum ranging from 430 nm to 1700 nm for full-frequency domain scanning. After passing through a series of optical components—including the linear polarizer, the focal lens with a focal length of 200 mm, and the beam splitter—the laser beam is focused onto the sample with an objective lens 10X. The transmitted light from the sample is split into two paths via a flip mirror: one directed to a CCD for imaging, and the other is collected by a spectrometer (AQ-6370D, Yokogawa). To control the polarization state of the incident light, a linear polarizer (GCL-051002, Daheng) in combination with a quarter-wave plate is employed. Notably, owing to the negligible absorption of silicon in the near-infrared regime, the measured normalized reflection and transmission spectra are nearly complementary.

**Nonlinear optical characterization.**

The nonlinear third-harmonic generation (THG) measurements is performed using the custom-built optical platform illustrated in Fig. S5. The excitation source comprised a femtosecond fiber laser (FemtoYL-100, YSL Photonics) coupled to an optical parametric amplifier (ParaAMP-8, Parametric Photonics), providing tunable near-infrared pulses over the wavelength range of 1400–1480 nm. The laser delivered pulses with a duration of approximately 200 fs at a repetition rate of ~200 kHz. The pump beam first passes through a variable attenuator (GCO-0703M, Daheng) and a Glan–Taylor prism (GCL-070223, Daheng) for precise control of the excitation power and polarization state. A quarter-wave plate is subsequently employed to generate the desired circularly polarized excitation. To suppress residual visible components from the laser source, a long-pass filter (MEFH10-700LP, LBTEK) is inserted into the optical path, ensuring that only the infrared pump beam reached the sample. The beam is then divided by a 50:50 non-polarizing beam splitter (GCC-M403132, Daheng), with one branch directed to a power meter for real-time monitoring of the incident pump power. The second branch is focused onto the membrane metasurface using a Mitutoyo M Plan Apo NIR 20× objective lens (NA = 0.40), producing a focal spot with a diameter of approximately 40 μm. The sample is mounted on a goniometric stage to control the angle of incidence of the pump light. The generated THG signal is collected in the reflection geometry using the same objective lens, as the reflection angles of the THG signal fell well within the numerical aperture of the objective lens. The reflected light is routed through a dichroic mirror (DM10-650LP, LBTEK), which efficiently rejected the residual fundamental pump while transmitting the visible-frequency THG emission. The resulting THG output is subsequently characterized either by a calibrated optical power meter (2936-R, Newport) for power measurements or by a fiber-coupled spectrometer (USB4000, Ocean Optics) for spectral analysis.

The nonlinear imaging configuration is derived directly from the THG measurement setup with only a minor modification. Specifically, a test target is inserted between the cold mirror and the dichroic mirror. In this configuration, the infrared image carried by the test target is projected onto the membrane metasurface and frequency upconverted through the THG process. The resulting visible image is then recorded by a sCMOS camera (pro.edge 5.5, PRO). The nonlinear imaging performance is evaluated using a Siemens star resolution target (RTR1CH-P, LBTEK) as well as customized “HFUT” patterns (abbreviation for Hefei University of Technology). Owing to the limited lateral dimensions of the fabricated metasurface samples, accurate sample alignment is essential. During the alignment procedure, an auxiliary LED illumination source is introduced to illuminate the sample surface. The reflected light is directed by a beam splitter to an imaging camera, enabling real-time monitoring of the spatial overlap between the focused beam and the metasurface area. Once optimal alignment isachieved, the LED illumination is switched off and the beam splitter is removed from the optical path during THG power and spectral measurements, thereby eliminating parasitic background signals and minimizing optical losses of the generated THG emission.

## Data availability

All data generated or analyzed during this study are included in the published article and its Supplementary Information. Additional data are available from the corresponding author on request.

## Acknowledgments

Baohe Zhang and Jumin Qiu contributed equally to this work.

## Funding

This work was supported by the National Natural Science Foundation of China (12574337, 12404359, 12304420, 12421005, 12247105, 12374273, and 12264028), the Anhui Provincial Natural Science Foundation (2308085MA24), the Hunan Provincial Major Sci-Tech Program (2023ZJ1010), and Young Elite Scientists Sponsorship Program by JXAST (2025QT04).

## Author contributions

H.L. conceived the idea and supervised the project. B.Z. performed numerical simulations and sample fabrication. J.Q. performed the linear and nonlinear optical characterization. B.Z. and H.L. analyzed the data and prepared the manuscript. S.X. and X. Z. co-supervised the project and participated in the numerical calculations. M.Q., H.J., Z.Z., L.K., and H.J. participated in the discussion. All authors participated in the manuscript writing.

## Competing interests

The authors declare no competing interests.

## Additional information

Supporting Information. The online version contains supplementary material available at.